\documentclass[twocolumn,showpacs,superscriptaddress,prl]{revtex4}
\usepackage{graphicx}
\usepackage{bm}

\begin{document}

\title{Time-dependent density-functional approach for exciton binding energies}

\author{V. Turkowski}
\affiliation{Department of Physics and Astronomy, University of Missouri, Columbia, MO 65211}
\affiliation{Department of Physics and NanoScience Technology Center, University of Central
Florida, Orlando, FL 32816}
\author{C. A. Ullrich}
\affiliation{Department of Physics and Astronomy, University of Missouri, Columbia, MO 65211}

\date{\today}

\begin{abstract}
Optical processes in insulators and semiconductors, including excitonic effects, can be described
in principle exactly using time-dependent density-functional theory (TDDFT). Starting
from a linearization of the TDDFT semiconductor Bloch equations in a two-band model,
we derive a simple formalism for calculating excitonic binding energies. This formalism leads to a
generalization of the standard Wannier equation for excitons, featuring a nonlocal effective
electron-hole interaction determined by long-range and dynamical exchange-correlation (XC) effects.
We calculate excitonic binding energies in several direct-gap semiconductors, using exchange-only and model XC kernels.
\end{abstract}


\maketitle

The elementary model  of Wannier excitons in
insulators views them as bound electron-hole pairs
which satisfy a hydrogen-like Schr\"odinger equation \cite{Marder2000}:
\begin{equation} \label{Wannier}
\left[-\frac{\hbar^2 \nabla^2}{2m_r} - \frac{e^2}{\epsilon r} \right]\phi({\bf r}) = E \phi({\bf r}) \:.
\end{equation}
Here, $m_r$ is the reduced electron-hole effective mass, $e$ is the electron charge, $\epsilon$ is the static dielectric
constant of the material, and $\phi$ and $E$ are the excitonic wave functions and binding energies
(from now on we set $\hbar=e=1$). Eq. (\ref{Wannier}),
also known as Wannier equation \cite{Wannier1937}, produces a Rydberg series of discrete energy states below
the conduction band edge, and a redistribution of oscillator strength in the optical spectrum
around the band edge which is qualitatively described by the Elliott formula \cite{Haug2004}.
Excitonic effects are important for a large variety of optical processes in organic and
inorganic materials and nanoscale systems \cite{Koch2006,Scholes2006}.

Eq. (\ref{Wannier}) can be derived from the semiconductor Bloch
equations within the time-dependent Hartree-Fock approximation,
using a dielectrically screened Coulomb interaction
\cite{Haug2004,Schafer2002}. It is well known that time-dependent
Hartree-Fock with bare Coulomb interaction leads to very poor
optical spectra of materials, with strongly overbound excitons. A
more rigorous ab-initio treatment of excitation processes in
insulators and semiconductors, including correlation-induced
screening, can be developed using many-body Green's function techniques such as
the GW/Bethe-Salpeter equation \cite{Onida2002}.

Time-dependent density-functional theory (TDDFT) \cite{TDDFT_book}
has recently emerged as an alternative, computationally convenient
approach to electronic excitation processes in materials
\cite{Onida2002,Reining,Kim,Botti2007}. In linear-response TDDFT,
excitation energies can be calculated in principle exactly \cite{Petersilka1996,Casida1996}, provided
the exchange-correlation (XC) kernel $f_{\rm xc}({\bf r},{\bf r}',\omega)$ is known. In
Refs. \cite{Onida2002,Reining}, an approximate $f_{\rm xc}$ was
constructed from many-body Green's functions, whereas Ref.
\cite{Kim} uses an exact-exchange (EXX) approach, including a cutoff in
wavevector space which mimics screening of the Coulomb interaction
\cite{Bruneval2006}. These studies have established that TDDFT is
capable of accurately describing excitonic effects in solids, although one needs
XC functionals that go beyond the more common ones such as
the adiabatic local-density approximation (ALDA) \cite{TDDFT_book}. The
resulting agreement with experimental data is excellent
\cite{Botti2007}, but the technical effort is not significantly less
than for standard many-body approaches.

The purpose of this paper is to develop a formally much simpler TDDFT treatment of excitonic effects in
solids. Rather than calculating complete optical spectra, our goal is more modest, namely a method that
directly yields excitonic binding energies, similar to the Wannier equation (\ref{Wannier}).
Starting from a TDDFT version of the semiconductor Bloch equations \cite{Turkowski2008},
we derive an effective electron-hole
interaction which explicitly shows how long-range XC effects are essential for exciton formation. Our
simplified treatment not only provides physical insight into the way excitonic effects
are treated in TDDFT, but also provides a straightforward way of testing approximate XC functionals.

{\em Time-dependent Kohn-Sham formalism for solids.} -- In TDDFT, the electron dynamics of a solid is described by
the time-dependent Kohn-Sham (KS) orbitals $\Psi_{j{\bf k}}({\bf r},t)$, where
$\bf k$ is the wave vector and $j$ is the valence band index (only the time evolution of the
initially occupied states is considered).
The system is assumed to start from the ground state, $\Psi_{j{\bf k}}({\bf r},t_0)=\psi_{j{\bf k}}({\bf r})$.
The KS Bloch functions and band structure follow from
\begin{equation}\label{Schroedinger}
\left(
 -\frac{\nabla^2}{2m}+V_{\rm lat}({\bf
r})+V_{\rm H}^0({\bf r}) +V_{\rm xc}^0({\bf r}) -\varepsilon_{j{\bf k}}
\right)\psi_{j{\bf k}}({\bf r})=0 \:,
\end{equation}
where $V_{\rm lat}$ is the crystal lattice potential (within the Born-Oppenheimer approximation) and $V_{\rm H}^0$ and
$V_{\rm xc}^0$ are the static Hartree and XC potentials.

Since the $\psi_{j{\bf k}}({\bf r})$ form a complete set for each $\bf k$,
we can expand the time-dependent KS orbitals as follows:
\begin{equation}\label{psi}
\Psi_{j{\bf k}}({\bf r},t)=\sum_l c_{\bf k}^{jl}(t)\psi_{l{\bf k}}({\bf r}) \:,
\end{equation}
where the summation runs over all valence and conduction bands, including continuum states.
Eq. (\ref{psi}) is appropriate if we assume the system to interact with an electromagnetic
field in dipole approximation. We define the density
matrix $\rho_{j{\bf k}}^{lm}(t)=c_{\bf k}^{jl}(t)[c_{\bf k}^{jm}(t)]^*$,
whose equation of motion is
\begin{equation}\label{Liouville}
i\frac{\partial}{\partial t}{\bm \rho}_{j{\bf k}}(t)
=[{\bf H}_{\bf k}(t),{\bm \rho}_{j{\bf k}}(t) ] ,
\end{equation}
with initial condition $\rho_{j{\bf k}}^{lm}(t_0)= \delta_{jl} \delta_{ml}$.
The matrix elements of the TDDFT Hamiltonian are
\begin{eqnarray} \label{Hamiltonianme}
H_{\bf k}^{lm}(t)&=&\frac{1}{\Omega}\int_{\Omega} d^3r \: \psi_{l{\bf k}}^*({\bf r}) H(t)
\psi_{m{\bf k}}({\bf r}) \nonumber \\
&=&
\varepsilon_{l{\bf k}}\delta_{lm}+{\bf E}(t){\bf
d}_{\bf k}^{lm} +\tilde{V}_{\rm H{\bf k}}^{lm}(t)+\tilde{V} _{\rm xc{\bf k}}^{lm}(t) \:,
\end{eqnarray}
where $\Omega$ is the volume of the lattice unit cell, ${\bf
E}(t)$ is the electric field amplitude, and ${\bf d}_{\bf k}^{lm}$
are the dipole matrix elements.
$\tilde{V}_{\rm H}(t) = V_{\rm H}(t)-V_{\rm H}^0$ denotes the
dynamic part of the Hartree potential, and similar for XC.
Self-consistent solution of Eq.~(\ref{Liouville}), with the time-dependent density
\begin{equation}
n({\bf r},t)=2 \sum_{j {\bf k}} \theta (\varepsilon_{F}-\varepsilon_{j{\bf k}})
\sum_{lm}
\rho_{j{\bf k}}^{lm}(t) \psi_{l{\bf k}}({\bf r}) \psi_{m{\bf k}}^*({\bf r}) \:, \label{nrho}
\end{equation}
where $\varepsilon_{F}$ is the Fermi energy,
is equivalent to solving the time-dependent KS equations for the solid, and is thus in principle exact.

{\em Two-band model and excitons}. -- To study optical excitation processes near the band gap, a two-band model
is a reasonable and widely used approximation. We consider one valence and one conduction band, $v$ and $c$,
assumed to be nondegenerate (see \cite{Baldereschi1971} for a discussion of band degeneracy).
The index $j$ of the density matrix $\rho_{j{\bf k}}^{lm}(t)$ refers to $v$ and will be
dropped in the following. Eq. (\ref{Liouville}) yields the TDDFT semiconductor Bloch equations for
the two independent components $\rho_{{\bf k}}^{vv}$ and $\rho_{{\bf  k}}^{vc}$ \cite{Turkowski2008}:
\begin{eqnarray}
\frac{\partial}{\partial t}\rho_{{\bf k}}^{vv}(t)
&=&
-2{\rm Im} \! \left\{ \left[{\bf E}(t){\bf d}_{\bf k}^{cv}+\tilde{V}_{H{\bf k}}^{cv}+\tilde{V}_{xc{\bf k}}^{cv} \right]
\rho_{{\bf k}}^{vc}(t)\right\}  \label{rhovv}
\\
i\frac{\partial}{\partial t}\rho_{{\bf k}}^{vc}(t)
&=&
\left[\varepsilon_{\bf k}^{v}-\varepsilon_{\bf k}^{c} +
\tilde{V}_{\rm H{\bf k}}^{vv}(t)+\tilde{V}_{\rm xc{\bf k}}^{vv}(t) - \tilde{V}_{\rm H{\bf k}}^{cc}(t) \right.
\nonumber\\
&& {}
\left.
-\tilde{V}_{\rm xc{\bf k}}^{cc}(t)  \right] \! \rho_{{\bf k}}^{vc}(t)
+ \left[{\bf E}(t){\bf d}_{\bf k}^{vc} + \tilde{V}_{\rm H{\bf k}}^{vc}(t) \right. \nonumber\\
&&
\left.
+ \tilde{V}_{\rm xc{\bf k}}^{vc}(t)\right] \! \left[\rho_{{\bf k}}^{cc}(t)-\rho_{{\bf k}}^{vv}(t) \right].
\label{rhovc}
\end{eqnarray}
Notice that
$\rho_{{\bf k}}^{vv}+\rho_{{\bf k}}^{cc}=1$ and $\rho_{{\bf  k}}^{vc}= \rho_{{\bf  k}}^{cv*}$.
In Ref. \cite{Turkowski2008}, Eqs. (\ref{rhovv}) and (\ref{rhovc}) were evaluated in the time domain
for ultrafast pulsed excitations.
Here, we are interested in excitonic binding energies, and we linearize Eq. (\ref{rhovc}):
\begin{equation} \label{rhovc_linearized}
i\frac{\partial}{\partial t}\rho_{{\bf k}}^{vc}(t) \!\!
=
[\varepsilon_{\bf k}^{v}-\varepsilon_{\bf k}^{c}] \rho_{{\bf k}}^{vc}(t)
- \delta\tilde{V}_{\rm H{\bf k}}^{vc}(t) - \delta\tilde{V}_{\rm xc{\bf k}}^{vc}(t) \:,
\end{equation}
where we dropped the time-dependent external field term, since the excitations we are interested in can be viewed as
eigenmodes of the system. Here, $\delta\tilde{V}_{\rm H{\bf k}}^{vc}$ and $\delta\tilde{V}_{\rm xc{\bf k}}^{vc}$
denote the linearized dynamical Hartree and XC potentials. In a periodic insulating solid, the Hartree term only
gives rise to the so-called local field corrections, which do not affect excitonic
binding \cite{Onida2002}. We will therefore only keep the XC contribution in the following.

Fourier transformation of Eq. (\ref{rhovc_linearized}) and the corresponding equation for $\rho_{{\bf k}}^{cv}(t)$
leads to
\begin{eqnarray} \label{smallmatrix1}
\rho_{\bf k}^{vc}(\omega) &=&
-\frac{\sum_{\bf q}
\left[F_{\bf kq}^{vccv}(\omega) \rho_{\bf q}^{vc}(\omega)
+ F_{\bf kq}^{vcvc}(\omega) \rho_{\bf q}^{cv}(\omega)\right]}{\omega + \omega_{\bf k}^{cv}}
\\
\rho_{\bf k}^{cv}(\omega) &=&
\frac{\sum_{\bf q}\left[F_{\bf kq}^{cvcv}(\omega) \rho_{\bf q}^{vc}(\omega)
+ F_{\bf kq}^{cvvc}(\omega) \rho_{\bf q}^{cv}(\omega)\right]}{\omega - \omega_{\bf k}^{cv}}
, \label{smallmatrix2}
\end{eqnarray}
where $\omega_{\bf k}^{cv} = \varepsilon_{\bf k}^c - \varepsilon_{\bf k}^v$,
\begin{eqnarray} \label{fxcmat}
F_{\bf kq}^{ijmn}(\omega) &=& \frac{2}{\Omega^2} \int_\Omega d^3r \!\int_\Omega
d^3r'\: \psi_{i{\bf k}}^*({\bf r})\psi_{j{\bf k}}({\bf r})f_{\rm
xc}({\bf r},{\bf r}',\omega)
\nonumber\\
&& \times
\psi_{m{\bf q}}^*({\bf r}')\psi_{n{\bf q}}({\bf r}') \:,
\end{eqnarray}
and the $\bf q$-summation
runs over the first Brillouin zone. Eqs. (\ref{smallmatrix1}) and
(\ref{smallmatrix2}) can be cast into an eigenvalue problem for the excitation energies $\omega$.
Since $f_{\rm xc}$ is in general frequency-dependent, the eigenvalue problem is nonlinear.
The solutions are the exact excitonic binding energies within the two-band model.

Let us carry out a further simplification. Since typical excitonic binding energies are much smaller than the band gap,
i.e., $\omega + \omega_{\bf k}^{cv} \gg \omega - \omega_{\bf k}^{cv}$, we can ignore the pole at negative $\omega$
(which is equivalent to the Tamm-Dancoff approximation \cite{TDDFT_book}) and boldly set $\rho_{\bf k}^{vc}=0$.
This leads to
\begin{equation} \label{singlepole}
\sum_{\bf q} \left[\omega_{\bf q}^{cv}\delta_{\bf kq} + F_{\bf kq}^{cvvc}(\omega)\right]\rho_{\bf q}^{cv}(\omega)
=\omega \rho_{\bf k}^{cv}(\omega) \:.
\end{equation}
Eq. (\ref{singlepole}) is the equivalent for extended systems of the well-known single-pole approximation
of linear-response TDDFT \cite{Petersilka1996}. For finite atomic or molecular systems,
the single-pole approximation only involves two discrete levels. Here, it involves two entire bands, which
clearly shows the collective nature of excitonic effects.

We point out that Eq. (\ref{singlepole}) yields excitonic binding energies relative to the conduction band edge,
which can be accurate even if the band gap itself is not.

{\em TDDFT Wannier equation.} -- Our next goal is to derive a real-space equation
for the excitonic binding energies. $\rho_{\bf k}^{cv}$ is a periodic function
in reciprocal space, with Fourier transform
$\rho({\bf R},\omega) = \sum_{\bf k} e^{-i\bf k\cdot R} \rho_{\bf k}^{cv}(\omega)$,
where ${\bf R}$ is a direct lattice vector.
Similarly, we define
\begin{eqnarray}\label{eh}
V_{eh}({\bf R},{\bf R}',\omega) &=& \sum_{\bf k,q} e^{-i\bf k \cdot R}
F_{\bf kq}^{cvvc}(\omega) e^{i\bf q \cdot R'} .
\end{eqnarray}
From the point of view of a Wannier exciton, which extends over many lattice constants, ${\bf R}$
can be approximated as a continuous variable. We assume
a direct band gap material, and use approximate parabolic
dispersions with conduction and valence band effective masses $m_c$
and $m_v$, and reduced electron-hole effective mass $m_r^{-1} =
m_c^{-1} + m_v^{-1}$.  This yields the TDDFT version of the Wannier equation (\ref{Wannier}),
\begin{equation} \label{Wannier-TDDFT}
\left[ -\frac{\nabla^2}{2m_r} - E_{b,i}\right]\rho_i({\bf r})
+ \int_{\rm all \atop space} \! d^3 r' V_{eh}({\bf r},{\bf r}',\omega) \rho_i({\bf r}')
=0 \:,
\end{equation}
featuring
a nonlocal, frequency-dependent electron-hole interaction $V_{eh}({\bf r},{\bf r}',\omega)$, where
$\omega = E_g^{\rm KS} + E_{b,i}$, and $E_g^{\rm KS}$ is the KS band gap.
The $i$th excitonic binding energy $E_{b,i}$ is measured with respect to the KS conduction band edge,
and the $\rho_i({\bf r})$ are the analog
of the excitonic wave functions $\phi({\bf r})$ of Eq. (\ref{Wannier}).

{\em XC kernels.} -- The effective electron-hole interaction in TDDFT,
and thus the excitonic binding energies, depend crucially on the approximate XC kernel. In the following,
we shall implement several simple frequency-independent XC kernels and test their performance in our formalism.

The exchange-only ALDA kernel is given by
\begin{equation}
f_{\rm x}^{\rm ALDA}({\bf r},{\bf r}')=-[9 \pi n_0^2({\bf r})]^{-1/3}\delta ({\bf r}-{\bf r'}) \:,
\label{fxcLDA}
\end{equation}
where $n_{0}({\bf r})$ is the equilibrium electron density. $f_{x}^{\rm ALDA}$ belongs to the class of ultra-short-range
kernels; the simplest of them is
\begin{equation}
f_{\rm xc}^{\rm contact}({\bf r},{\bf r}') = -A \delta ({\bf r}-{\bf r'}) \:,
\label{fxcContact}
\end{equation}
where $A$ is a positive constant. Such kernels have been used with some success in contact exciton models \cite{Botti2007}.

An approximation of exact-exchange TDDFT \cite{Kim}, the Slater exchange kernel, is given by \cite{Petersilka1996}
\begin{eqnarray}
f_{\rm x}^{\rm Slater}({\bf r},{\bf r}')=-\frac{2|\sum_{j{\bf k}}\theta(\epsilon_F - \epsilon_{j{\bf k}})
\psi_{j \bf k}({\bf r})\psi_{j\bf k}^{*}({\bf r}')|^{2}}{|{\bf
r}-{\bf r}'|n_{0}({\bf r})n_{0}({\bf r}')}.
 \label{fxcSlater}
\end{eqnarray}
This kernel exhibits some degree of long-range behavior  \cite{Lein2000}, but not the ultra-nonlocality
($\sim 1/q^2$ in momentum space) of the exact $f_{\rm xc}$ \cite{Onida2002,Kim,Botti2007}.
This long-range contribution (LRC) can be explicitly taken into account using the following model kernel \cite{Botti2004}:
\begin{equation}
f_{\rm xc}^{\rm LRC}({\bf r},{\bf r}') = -\frac{\alpha}{4 \pi |{\bf r}-{\bf r'}|} \:,
\label{fxcLR}
\end{equation}
where $\alpha$ is again an adjustable parameter.

\begin{table}
\caption{ $E_{b}^{\rm exp}$ and $E_{b}^{\rm Slater}$: lowest direct excitonic binding energies (in meV)
for selected III-V and II-VI
compounds, from experiment \cite{Landolt} and from Eqs. (\ref{singlepole}), (\ref{fxcmat}) with $f_{\rm x}^{\rm Slater}$.
The parameters $A$ and $\alpha/4\pi$ (in a.u.) are fitted to reproduce $E_{b}^{\rm exp}$
using Eqs. (\ref{singlepole}), (\ref{fxcmat}) with $f_{\rm xc}^{\rm contact}$ and $f_{\rm xc}^{\rm LRC}$.
}
\begin{ruledtabular}
\begin{tabular}{ccccc}
            & $A$      &  $\alpha /4\pi$  &     $E_{b}^{\rm Slater}$  &  $E_{b}^{\rm exp}$  \\
GaAs        & 1.68     &  0.12            &  17.4        &  3.27        \\
$\beta$-GaN & 4.23     &  0.55            &  28.7        &  26.0        \\
$\alpha$-GaN& 2.03     &  0.91            &  11.8        &  20.4        \\
CdS         & 6.28     &  1.83            &   7.9        &  28.0        \\
CdSe        & 4.84     &  1.19            &   8.3        &  15.0
\end{tabular} \label{table1}
\end{ruledtabular}
\end{table}

{\em Results and Discussion.} -- We have tested our TDDFT approach for excitonic binding energies,
Eq. (\ref{singlepole}), for
the zincblende materials GaAs and $\beta$-GaN and for the wurtzite materials $\alpha$-GaN, CdS, and CdSe.
The Bloch functions for the conduction and heavy-hole valence bands were obtained
from LDA band structures calculated with the plane-wave pseudopotential code ABINIT \cite{ABINIT}.
We used 512 $\bf k$-points in the first Brillouin zone for all materials. Out of these, there are 10 independent
points for GaAs and $\beta$-GaN and 20  for $\alpha$-GaN, CdS, and CdSe,
which determines the dimension of the eigenvalue problem (\ref{singlepole}). Recent
Bethe-Salpeter calculations of excitonic binding energies used
much higher $\bf k$-point densities close to the zone center \cite{Laskowski2005,Fuchs2008};
we performed convergence checks of our $\bf k$-point sampling rates and found them to be sufficiently accurate
for our simple model.

As expected, the ALDA does not produce any bound excitons.
Results for the other three XC kernels and experimental binding energies of the lowest direct excitons
are presented in Table \ref{table1}. The contact and LRC kernels, (\ref{fxcContact}) and (\ref{fxcLR}),
contain adjustable parameters which can be tuned to reproduce the experimental exciton binding energies.
The required values of the parameters $A$ and $\alpha$ are found to be of similar order as in Ref. \cite{Botti2004}.

The contact and LRC kernels only yield a single excitonic bound state \cite{Botti2004}.
This is generally the case for static XC kernels that are local in reciprocal space, i.e., have
the form $f_{\rm xc}({\bf q})$.
The Slater XC kernel (\ref{fxcSlater}) does have some degree of nonlocality in reciprocal space, but we found
that it only produces a single excitonic state, like the local kernels.
To obtain an excitonic Rydberg series one needs an XC kernel that
has a sufficiently strongly nonlocal form or is frequency-dependent \cite{Sottile2003,Sottile2007}.

Looking at the results obtained with $f_{\rm x}^{\rm Slater}$, we find
excitons that are overbound by 14 meV in GaAs and by 2.7 meV in $\beta$-GaN.
This overbinding is what one would expect from an unscreened exchange-only approach (electronic screening
can be viewed as a correlation effect).
On the other hand, $f_{\rm x}^{\rm Slater}$ approaches a constant
for $q\to 0$ in homogeneous systems \cite{Lein2000}, whereas the full EXX $f_{\rm x}$ behaves as $1 / q^2$ \cite{Kim}.
This would suggest that $f_{\rm x}^{\rm Slater}$ has a somewhat weaker
effective electron-hole interaction than full EXX. This trend seems confirmed in the wurtzite materials
whose calculated excitonic binding energies are significantly below experiment.

\begin{figure}
\centering
\includegraphics[origin=b,angle=0,width=0.9\linewidth]{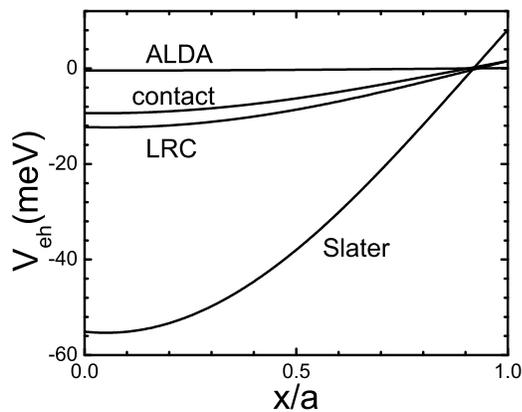}
\caption{Effective electron-hole interaction $V_{eh}({\bf r},0)$ for GaAs and different XC kernels, plotted along the $x$
direction, where $a$ is the lattice constant. The parameters $A$ and $\alpha$  for the contact and the LRC XC kernels
are given in Table \ref{table1}. } \label{fig1}
\end{figure}

Additional insight is provided by comparing the electron-hole interaction $V_{eh}$
for the different XC kernels under study. Fig. \ref{fig1} shows $V_{eh}({\bf r},0)$ for GaAs along the
$x$ direction (due to the finite sampling in $\bf k$-space, $V_{eh}$ can only be reliably calculated within the range
of about one unit cell). In ALDA, the interaction is close to zero and thus too shallow to lead to any
excitonic binding. The other XC kernels produce stronger electron-hole interactions, where for GaAs
the contact and LRC models are less attractive than the Slater approximation.

{\em Conclusion.} --
We have presented a simple method to calculate excitonic binding energies using TDDFT.
The main idea, restricting the dynamics to the highest valence and the lowest conduction band,
is similar to the single-pole approximation for excitation energies \cite{Petersilka1996}. Our derivation
was based on the TDDFT semiconductor Bloch equation; an alternative starting point could be the Casida formalism of
linear-response TDDFT \cite{Casida1996}, formulated for periodic systems \cite{Gruning2007}.
The resulting simple eigenvalue equation in momentum space, Eq. (\ref{singlepole}), is readily diagonalized
to yield the excitonic binding energies. Transformation into real space leads to the TDDFT analog of the
Wannier equation for excitons, and shows that the effective electron-hole interaction is nonlocal.

The quality of the results depends crucially on the approximation used for $f_{\rm xc}({\bf r},{\bf r}',\omega)$.
It is well known that local and semilocal approximations, such as the ALDA, do not produce any excitons.
There exist sophisticated parameter-free XC kernels \cite{Reining} that are capable of
reproducing experimental optical absorption spectra very accurately, including bound excitons \cite{Sottile2007},
but with substantial computational cost.

If only particular aspects of the optical spectrum of a material are required such as,
for instance, the lowest bound exciton, simple static XC kernels can be a convenient alternative.
The contact and the LRC kernels behave quite similarly in the sense that they
produce a single excitonic peak. A detailed analysis was given in Ref. \cite{Sottile2003}, and we
find the same behavior in our two-band approach. The parameter-free Slater exchange-only kernel also produces a single exciton,
which was found to be overbound in zincblende materials, and underbound in wurtzite. There are theoretical
arguments in favor of both  trends, which suggests a need for more systematic studies of the Slater exchange kernel in
solids.

In conclusion, our simple approach for excitonic binding energies is a promising method to test XC kernels in solids.
It can be extended in a straightforward way to deal with spin-dependent excitations (triplet excitons), with
more sophisticated XC kernels, or to include more bands. Work along these lines is in progress.

\acknowledgments
This work was supported by NSF Grant DMR-0553485. We thank Angel Rubio, Lucia Reining,  and Claudia Ambrosch-Draxl for
useful discussions.

\end{document}